# Superstatistics in random matrix theory

A.Y. Abul-Magd

*Department of Mathematics, Faculty of Science,*
*Zagazig University, Zagazig, Egypt*

**Abstract**

Using the superstatistics method, we propose an extension of the random matrix theory to cover systems with mixed regular-chaotic dynamics. Unlike most of the other works in this direction, the ensembles of the proposed approach are basis invariant but the matrix elements are not statistically independent. Spectral characteristics of the mixed systems are expressed by averaging the corresponding quantities in the standard random-matrix theory over the fluctuations of the inverse variance of the matrix elements. We obtain analytical expressions for the level density and the nearest-neighbor-spacing distributions for four different inverse-variance distributions. The resulting expressions agree with each others for small departures from chaos, measured by an effective non-extensivity parameter. Our results suggest, among other things, that superstatistics is suited only for the initial stage of transition from chaos to regularity.

*Keywords*: Superstatistics, Non-extensive statistics, Random-matrix theory, Mixed systems.

## 1. Introduction

According to a conjecture by Bohigas, Giannoni and Schmidt [1], the quantum spectra of classically chaotic systems are correlated according to the random matrix theory (RMT) [2], which models the Hamiltonian by an ensemble of basis-independent matrices of uncorrelated elements, whereas the spectral correlations of classically integrable systems are close to Poissonian statistics. This conjecture has been checked numerically on a wide variety of systems (for recent reviews, see e.g. [3]). Completely regular or chaotic classical systems are rare. For most systems, the phase space is partitioned into regular and chaotic regions. These systems are referred to as mixed systems. It is therefore important to find models of random matrices that describe the spectral fluctuations of such mixed systems.

Rosenzweig and Porter [4] were probably the first to formulate a RMT for mixed systems. Their model is governed by a Hamiltonian, which is essentially a sum of two terms, one for the chaotic part of the phase space and one for the regular. It has been elaborated on and further developed by several authors (for reviews see, e.g. [3,5]). Ensembles of such Hamiltonian matrices are often referred to as the deformed Gaussian ensembles [6]. Another extension of RMT along this line is achieved by introducing band matrices in which the variances of the Hamiltonian matrix element decrease with the distance from the diagonal. The Hamiltonian matrix elements in these models are independent Gaussian variables, with different variances for the diagonal and off-diagonal elements, so that the departure from chaos is expressed by violating the basis invariance.

Another possibility for describing mixed system by RMT is to conserve basis invariance but violate matrix-element independence. One way to do this is to start the statistical treatment from Tsallis' non-extensive entropy [7] instead of Shannon's entropy that leads to the Gaussian random ensembles [8]. Then the departure from chaos towards order is spelled out by varying Tsallis' entropic index from 1 to a certain upper limit that depends on the dimension of the ensemble [9] and the symmetry class of the Hamiltonian [10]. Beyond this limit, the integrals involved in the theory diverge. Therefore, the resulting non-extensive RMT can only describe mixed systems not quite far from chaos. It is not able to describe nearly regular systems whose dynamics follows Poisson statistics. Eigenfunctions of an integrable system form a basis, which corresponds in the classical description to the invariant tori. In the light of the well-known Kolmogorov-Arnold-Moser theorem, one expects this basis to play a special role in quantum description when the system is slightly perturbed. Its role fades out as the system moves further towards chaos.

The present paper develops an analogous basis invariant RMT of mixed systems which is based on superstatistics, recently introduced by Beck and Cohen [11], which includes Tsallis' non-extensive statistics as a special case. Superstatistics has been successfully applied to a wide variety of physical problems, including turbulence [12], plasma physics [13], cosmic-ray statistics [14], and econophysics [15]. Here we introduce the notion of superstatistics into RMT. Beck and Cohen expressed the distribution function for a non-equilibrium state of a system as a mixture of equilibrium (Gibbs) distributions characterized by a fluctuating inverse temperature. We represent the matrix-element distribution for a mixed system as a superposition of Gaussian distributions that model chaotic systems by allowing a fluctuation of their characteristic parameter, namely the inverse variance of the matrix elements. We are then able to express the characteristics of systems evolving out of chaos in terms of the corresponding characteristics of chaotic systems obtained by the standard RMT. We obtain closed-form expressions for the level density and spacing distributions for four different parameter distributions.

## 2. Superstatistics for mixed systems

Ballian [8] derives the matrix-element distribution functions in RMT by applying the principle of maximum entropy. Assuming Shannon's entropy, and constraining the expectation value of $\text{Tr}(\mathbf{H}^+\mathbf{H})$ for an ensemble of random matrices $\mathbf{H}$ to take a constant value,

$$\int d\mathbf{H}\,\text{Tr}(\mathbf{H}^+\mathbf{H})\exp[-\eta\text{Tr}(\mathbf{H}^+\mathbf{H})] = \text{const}, \qquad (1)$$

he obtains a Gaussian distribution function

$$P^{(G)}(\mathbf{H}) = \frac{1}{Z(\eta)}\exp[-\eta\text{Tr}(\mathbf{H}^+\mathbf{H})], \qquad (2)$$

where $Z(\eta) = (\eta/\pi)^{N/2+N(N-1)\beta/4}$ is a normalization constant, $N$ the dimension of the matrix $\mathbf{H}$ and $\eta$ a Lagrange multiplier associated with the constraint (1). Here $\beta = 1, 2$ and 4 for the Gaussian orthogonal (GOE), unitary (GUE) and symplectic (GSE) ensembles, respectively. Thus for a GOE, where $\text{Tr}(\mathbf{H}^+\mathbf{H}) = \sum_i H_{ii}^2 + 2\sum_{i<j} H_{ij}^2$, $\eta = 1/4a^2$ with $a$ being the dispersion of the nondiagonal matrix elements.

We introduce superstatistics into RMT by constraining the expectation value of Tr($\mathbf{H}^+\mathbf{H}$) defined by Eq. (1) to be a random number rather than a definite one. The model parameter η will be no longer fixed but characterized by a distribution of $f(\eta)$. The distribution function for mixed systems is given by a superposition of two statistics, that of η and that of Tr($\mathbf{H}^+\mathbf{H}$)

$$P(\mathbf{H}) = \int_0^\infty d\eta \frac{1}{Z(\eta)} \exp\left[-\eta \text{Tr}(\mathbf{H}^+\mathbf{H})\right] f(\eta). \quad (3)$$

In the absence of fluctuation of the inverse variance, we set $f(\eta) = \delta(\eta-\eta_0)$ and obtain the distribution function (2) of the conventional RMT. Hence, the proposed superstatistics formalism associates the departure from chaos with the fluctuation of the inverse variance of matrix elements. Taking $f(\eta)$ to be a $\chi^2$-distribution yields Tsallis' statistics as shown in [16, 17]. The resulting distribution functions, studied in [9, 10], are possible candidates for describing mixed systems. There are infinite other possibilities for the choice of $f(\eta)$; some of them are considered by Beck and Cohen [11]. We shall return to this point later.

The distribution function of the superstatistical ensemble, defined by Eq. (3), depends on the Hamiltonian matrix $\mathbf{H}$ through Tr($\mathbf{H}^+\mathbf{H}$). We note that two matrices $\mathbf{A}$ and $\mathbf{B}$, which express the same operator in two different bases are related by a similarity transformation $\mathbf{B} = \mathbf{P}^{-1}\mathbf{AP}$. Such operators have the same trace. Therefore, the proposed superstatistical RMT is basis independent as the standard RMT. However, Eq. (3) does not allow the factorization of $P(\mathbf{H})$ into products of terms depending on individual matrix elements as the distribution function in Eq. (2). The matrix elements can no more be considered as independent random numbers.

The eigenvalue distribution is obtained by introducing the diagonal matrix $\mathbf{E} = \text{diag}(E_1, ..., E_N)$ by means of the unitary transformation $\mathbf{H} = \mathbf{U}^{-1}\mathbf{EU}$ [2, 3]. Taking the eigenvalues and independent elements of $\mathbf{U}$ as new variables, the Cartesian volume of $\mathbf{H}$ is given by

$$d\mathbf{H} = \left[\prod_{n>m}(E_n - E_m)\right]^\beta d\mathbf{E}d\mu(\mathbf{U}), \quad (4)$$

where $d\mu(\mathbf{U})$ is the Haar measure of the group $\mathbf{U}$. Integrating over the "angular" variables $\mathbf{U}$, and noting that Tr($\mathbf{H}^+\mathbf{H}$) is invariant under rotation in the matrix-element space, we obtain the eigenvalues distribution

$$P_\beta(\mathbf{E}) = \int_0^\infty d\eta P_\beta^{(G)}(\eta, \mathbf{E}) f(\eta), \quad (5)$$

where

$$P_\beta^{(G)}(\eta, \mathbf{E}) = C_\beta \eta^{N/2+\beta N(N-1)/4}\left[\prod_{n>m}(E_n - E_m)\right]^\beta \exp\left(-\eta \sum_{n=1}^N E_n^2\right). \quad (6)$$

Here

$$C_\beta = 2^{\beta N(N-1)/4} \pi^{-N/2} \Gamma^N(1+\beta/2) \Big/ \prod_{n=1}^N \Gamma(1+\beta n/2),$$

where $\Gamma(x)$ is Euler's gamma function. We thus obtain

$$P_\beta(\mathbf{E}) = C_\beta\left[\prod_{n>m}(E_n - E_m)\right]^\beta \int_0^\infty d\eta \eta^{N/2+\beta N(N-1)/4} \exp\left(-\eta \sum_{n=1}^N E_n^2\right) f(\eta), \quad (7)$$

We note that the level repulsion in the joint eigenvalue distribution is the same for every choice of the parameter distribution $f(\eta)$. Equation (7) cannot produce the Poisson level spacing distribution, which is characteristic for nearly integrable systems. Therefore, the proposed random matrix ensemble is not suited for describing the whole chaotic-integrable transition. It may be useful only when the system is not far from chaos. We have already seen this while applying Tsallis' non-extensive statistics to RMT [10, 11]. There, the convergence of the normalization integrals, and the ones involved in the constraint in Eq. (1), puts an upper limit on the entropic index (see Subsection 3.1 below). The spacing distribution corresponding to this limit has a shape intermediate between the Wigner and Poisson distributions. As the mixed system evolves further towards integrability, the proposed description may not be valid because of possible violation of basis invariance. One expects a gradual establishment of approximately conserved quantum number. Models in which the spectrum of the mixed system consists of a superposition of RMT spectra, as e.g. in [18, 19], may be more suitable for this stage of the stochastic transition.

Several commonly used spectral characteristics can be, in principle, obtained from the eigenvalue joint probability distribution function (7) by integration over some of its variables [2, 3]. Among them are the level density, the nearest-neighbor spacing distribution and the higher-order ones as well as the two-level correlation functions from which one derives the level-number variance $\Sigma^2$ and the level rigidity $\Delta_3$. Superstatistics presents a recipe for evaluating these characteristics for mixed systems. Firstly, express the corresponding characteristic for an analogous chaotic system explicitly in terms of the parameter $\eta$ of the matrix element distribution, and secondly, take the average with respect to its probability density function $f(\eta)$. The remaining task is to define the function $f(\eta)$.

## 3. Inverse-variance distribution

Beck and Cohen [11] gave some examples of functions which are potential candidates for the inverse-variance distribution $f(\eta)$ of the Hamiltonian matrix elements. They found in considering non-equilibrium thermodynamical systems that deviations from the Gibbs distribution due to small variance of the inverse-temperature fluctuations are insensitive to the choice of the inverse-temperature distribution. We shall show that a similar insensitivity to the form of $f(\eta)$ holds in RMT at least when the system under consideration is not far from the state of chaos. In the following we consider only some of their proposals, namely those for which we can do the calculations analytically.

### 3.1. Gamma distribution

The choice of $f(\eta)$ as a gamma (or $\chi^2$-) distribution

$$f(\eta) = \frac{1}{\Gamma(n/2)} \left(\frac{\eta}{2\eta_0}\right)^{\frac{n}{2}} \eta^{\frac{n}{2}-1} e^{-\frac{n\eta}{2\eta_0}}, \qquad (8)$$

which is the distribution of the sum of squares of $n$ Gaussian random numbers of zero mean, yields the Tsallis random-matrix ensembles [10] in which the entropic index $q = 1 + 2/(n + 1)$. This has been shown in the general case by Wilk and Wlodarczyk [16] and Beck [17]. It has a mean value $\langle\eta\rangle = \eta_0$ and a second moment

$$\langle\eta^2\rangle = q\langle\eta\rangle^2. \qquad (9)$$

This relation will be used to define an effective non-extensivity parameters for the other choices of $f(\eta)$. It is shown in [10] that the normalization condition for the matrix-element distribution together with the constraint of a fixed value of $\text{Tr}(\mathbf{H}^+\mathbf{H})$ impose an upper limit for the admissible values of $q$. For a two-dimensional GOE

$$q \leq 1 + \frac{2}{\beta+2} = \frac{7}{5} . \qquad (10)$$

Increasing the dimension of the ensemble reduces the upper limit of $q$ as shown by Toscano et al. [9].

*3.2. Uniform distribution*

We start by a uniform distribution of the inverse matrix-element variance $\eta$ in an interval of width $b$ centered at $\eta_0$, i.e.

$$f(\eta) = \begin{cases} \frac{1}{b}, & \text{for } 0 < \eta_0 - b/2 < \eta < \eta_0 + b/2 \\ 0 & \text{elsewhere} \end{cases}, \qquad (11)$$

where $b < 2\eta_0$. With this distribution the eigenvalue distribution (7) converges to that of the standard RMT in the limit of $b = 0$. It has a mean of $\eta_0$ and a variance of $b^2/12$. We use Eq. (9) to define for it (and for the other distribution to be considered below) an effective non-extensivity parameter as

$$q = 1 + \frac{b^2}{12\eta_0^2} \leq \frac{4}{3} , \qquad (12)$$

where the inequality is required to prevent negative values of $\eta$. Thus the origin of upper bounds on $q$ given in (10) and (12) are different.

*3.3. Two-level distribution*

This distribution describes the situation where the sub-spectra can switch between two discrete values of the local mean level density. The parameter-probability density distribution is given as a sum of Dirac's delta-functions by

$$f(\eta) = \frac{1}{2}\delta(\eta - \eta_0 + b/2) + \frac{1}{2}\delta(\eta - \eta_0 - b/2). \qquad (13)$$

Here, $b < 2\eta_0$ in order to exclude the possibility of negative $\eta$. The mean value of this distribution is again $\eta_0$ while its variance is of $b^2/4$. The corresponding effective non-extensivity is given by

$$q = 1 + \frac{b^2}{4\eta_0^2} \leq 2 . \qquad (14)$$

*3.4. Gaussian distribution*

Beck and Cohen [11] pointed out that Gaussian parameter fluctuations would generate one of the simplest superstatistics. They did not consider Gaussian parameter distributions since these allow for negative inverse temperatures with non-zero probability. We shall amend this situation by considering the following modified Gaussian distribution

$$f(\eta) = \frac{1}{b\sqrt{\pi}\,\text{erf}(\eta_0/b)} \left[ e^{-\left(\frac{\eta-\eta_0}{b}\right)^2} - e^{-\left(\frac{\eta+\eta_0}{b}\right)^2} \right] \qquad (15)$$

for positive values of $\eta$, whereas $f(\eta) = 0$ for $\eta \leq 0$. The pre-factor takes care of the normalization condition. The mean value of this distribution is $\eta_0$. Dividing the second moment for this distribution by the square of the first one yields, according to Eq. (9), an effective non-extensivity index given by

$$q = 1 + \frac{1}{2}\left(\frac{b}{\eta_0}\right)^2 + \frac{b\exp\left[-(\eta_0/b)^2\right]}{\eta_0\sqrt{\pi}\operatorname{erf}(\eta_0/b)}. \tag{16}$$

## 4. Level Density

The eigenvalue density can be obtained by integrating the joint eigenvalue distribution over all variables except one. Using Eq. (5), we obtain for a superstatistical ensemble

$$\rho_\beta(E) = \int_0^\infty d\eta \, \rho_\beta^{(G)}(\eta, E) f(\eta), \tag{17}$$

where $\rho_\beta^{(G)}(\eta, E)$ is the level density for a Gaussian ensemble with parameter η. The latter density is given, for large $N$, by Wigner's semi-circle law (see [20], p. 60, and change the notation $4a^2$ into $1/\eta$)

$$\rho_\beta^{(G)}(\eta, E) = \frac{2}{\pi}\sqrt{\frac{\eta N}{\beta}\left(1 - \frac{\eta E^2}{\beta N}\right)}\,\Theta\!\left(1 - \frac{\eta E^2}{\beta N}\right), \tag{18}$$

where $\Theta(x)$ is the Heaviside step function. Substituting (18) into (17), we obtain

$$\rho_\beta(E) = \frac{2}{\pi}\sqrt{\frac{N}{\beta}}\int_0^{\beta N/E^2}\sqrt{\eta\left(1 - \frac{\eta E^2}{\beta N}\right)}f(\eta)\,d\eta. \tag{19}$$

We have used this relation to evaluate the level density distribution for the four superstatistical random-matrix ensembles that correspond to the parameter distributions defined in the preceding section by Eqs. (8), (11), (13) and (15). We confine our consideration to the cases where time reversal invariance holds, i.e. when β = 1. We use Mathematica software [21] to evaluate the integrals, which becomes possible if one rewrites Eq. (19) in the following form

$$\rho_\beta(E) = \frac{2}{\pi}\sqrt{\frac{N}{\beta}}\operatorname{Re}\!\left[\int_0^\infty \sqrt{\eta\left(1 - \frac{\eta E^2}{\beta N}\right)}f(\eta)\,d\eta\right], \tag{20}$$

where Re[$w$] stand for the real part of $w$. The change of the upper bound in Eq. (19) to infinity in Eq. (20) is possible because $\operatorname{Re}\sqrt{1 - \eta E^2/\beta N} = 0$ for $\eta > \beta N/E^2$. We then obtain in the case when $f(\eta)$ is given by a Gamma distribution in Eq. (8)

$$\rho_{Ts}(E) = \frac{1}{\sqrt{\pi}}\left(\frac{nN}{2\eta_q}\right)^{n/2}|E|^{-1-n}\frac{\Gamma\!\left(\frac{1+n}{2}\right)}{\Gamma\!\left(2+\frac{n}{2}\right)\Gamma\!\left(\frac{n}{2}\right)}\,{}_1F_1\!\left(\frac{1+n}{2}, 2+\frac{n}{2}, -\frac{nN}{2\eta_q E^2}\right), \tag{21}$$

where ${}_1F_1(a,b,x)$ is the confluent hypergeometric function [22]. In the case of the uniform distribution (9), we obtain

$$\rho_{\text{Uniform}}(E) = \frac{2N}{\pi b|E|^3}\operatorname{Re}\!\left[\Phi\!\left(\frac{E^2(\eta_0 - b/2)}{N}, \frac{E^2(\eta_0 + b/2)}{N}\right)\right], \tag{22}$$

where

$$\Phi(x, y) = \frac{1}{8}\left[2\sqrt{y(1-y)}(2y-1) - 2\sqrt{x(1-x)}(2x-1) + \arcsin(2y-1) - \arcsin(2x-1)\right].$$

If $f(\eta)$ has a two-level distribution as given by Eq. (10). Then the level density is given by

$$\rho_{\text{two-level}}(E) = \frac{1}{\pi\sqrt{N}} \text{Re}\left[\sqrt{\left(\eta_0 - \frac{b}{2}\right)\left\{1-\left(\eta_0-\frac{b}{2}\right)\frac{E^2}{N}\right\}} + \sqrt{\left(\eta_0 + \frac{b}{2}\right)\left\{1-\left(\eta_0+\frac{b}{2}\right)\frac{E^2}{N}\right\}}\right].$$

(23)

Unfortunately, we have not been able to obtain an analytic expression for the level density in the case of the modified Gaussian distribution in Eq. (10).

We choose an energy scale that makes the central level density of the Gaussian ensemble $\rho_\beta^{(G)}(\eta_0, 0) = 1$. For this purpose, we take $\eta_0 = (\pi/2)^2 \beta/N$. Furthermore, we use Eqs. (12), (14) and (16) to express the parameters $b$ of the uniform, two-level and modified Gaussian distributions in terms of the effective parameter $q$. With these values, we calculate the level densities given by Eqs. (21)-(23), and numerical integrate Eq. (20) for the case of a modified Gaussian distribution. The result of calculation is shown in Fig. 1 for effective $q = 1, 1.1$ and $1.3$. The figure shows that, in all of the four cases, the level density is symmetric with respect to $E = 0$ for all values of $q$ and has a pronounced peak at the origin. However, the behavior of the level density for $q > 1$ is quite distinct from the semicircular law. It has a long tail whose shape and decay rate both depend on the choice the parameter distribution $f(\eta)$. Introducing a new variable $x = 1 - \eta E^2/\beta N$ in the integration of the right-hand-side of Eq. (19) one can easily see that the asymptotic behavior of the level density is dominated by a factor of $|E|^{-3}$.

We note that near the origin, one can approximately replace the upper limit of the integral in (19) by infinity, and expand the square root in the integrand in powers of $E$ to obtain, in particular,

$$\rho_\beta(0) = \rho_\beta^{(G)}(\eta_0, 0) \frac{\langle\sqrt{\eta}\rangle}{\sqrt{\langle\eta\rangle}}.$$

(24)

Explicitly, we have

$$\frac{\rho_\beta(0)}{\rho_\beta^{(G)}(\eta_0,0)} = \sqrt{\frac{2}{n}}\frac{\Gamma((1+n)/2)}{\Gamma(n/2)}, \quad \text{(Gamma)}$$

$$= \frac{2\eta_0}{3b}\left[\left(1+\frac{b}{2\eta_0}\right)^{3/2} - \left(1-\frac{b}{2\eta_0}\right)^{3/2}\right], \quad \text{(uniform)}$$

$$= \frac{1}{2}\left(\sqrt{1+\frac{b}{2\eta_0}} + \sqrt{1-\frac{b}{2\eta_0}}\right), \quad \text{(2-level)}$$

$$= \frac{2\Gamma(4)\sqrt{\eta_0}}{\sqrt{\pi}b\,\text{Erf}\left(\frac{\eta_0}{b}\right)} e^{-\left(\frac{\eta_0}{b}\right)^2} {}_1F_1\left(\frac{5}{4},\frac{3}{2},\frac{\eta_0^2}{b^2}\right), \quad \text{(Gaussian)}$$

(25)

The dependence of the central level density $\rho_\beta(0)$ on the effective non-extensivity index $q$ is demonstrated in Fig. 2. Both Figs. 1 and 2 show that different parameter distribution corresponding to equal values of $q$ yield similar level density distributions, which differ only by few percents. The difference is largest at the peripheral values of *E* which are less interesting, because most of the spectral fluctuation analysis usually disregard levels at the beginning and end of the spectrum.

**5. Nearest-Neighbor-Spacing Distribution**

RMT does not provide simple analytical expressions for the nearest neighbor spacing (NNS) distributions. There are several elaborate approaches to evaluate these distributions. For example, Mehta [2] expresses the gap function $E_\beta(s)$ ($p_\beta(s) = d^2 E_\beta(s)/ds^2$) as an infinite product of the factors $[1 - \lambda_n(s)]$, where $\lambda_n(s)$ are eigenvalues of certain integral equations. Another approach, which has recently been the subject of numerous investigations, is based on the relation to differential equations of the Painlevé type [23]. It expresses the gap function $E_\beta(s)$ in terms of integrals involving a function $\sigma(t)$ which satisfies a nonlinear second-order differential equation. These approaches result in tabulated numerical values, series expansions and asymptotic expressions for the NNS distributions. The author is not aware of application of these results to the analysis of experimentally observed or numerically calculated discrete data.

In many cases, the empirical data are actually compared to a so-called Wigner surmise, which corresponds to the NNS distributions of ensembles of 2×2 matrices. For the three Gaussian ensembles, the Wigner surmise is given by

$$p_\beta(s) = a_\beta s^\beta \exp(-b_\beta s^2), \qquad (26)$$

where $(a_\beta, b_\beta) = (\pi/2, \pi/4)$, $(32/\pi^2, 4/\pi)$, and $(2^{18}/3^6\pi^3, 64/9\pi)$ for GOE, GUE, and GSE, respectively. They present accurate approximation to the exact results for the case of large *N*. Ensembles of 2×2 matrices have also been successfully used to model mixed systems by several authors (see, e.g., [24, 25] and references therein).

We consider the superstatistical generalization of the Wigner surmise hoping that it presents an accurate approximation for superstatistical ensembles of large *N*, which has an equal success to that of the standard Wigner surmise. For this purpose, we apply Eq. (5) to the case of *N* = 2. Then, the joint eigenvalue probability density function $P_\beta^{(G)}(\eta, E_1, E_2)$ is given by Eq. (6) with *N* = 2. Introducing the new variable $s = |E_1 - E_2|$ and $E = (E_2 + E_1)/2$ into Eq. (5) and integrating over *E*, we obtain the following expression for the NNS distribution

$$P_\beta(s) = \frac{1}{2^{\frac{\beta-1}{2}} \Gamma\left(\frac{\beta+1}{2}\right)} s^\beta \int_0^\infty d\eta\, \eta^{\frac{\beta+1}{2}} f(\eta) e^{-\frac{1}{2}\eta s^2}. \qquad (27)$$

The condition of unit mean spacing implies changing the energy scale in the definition of $f(\eta)$ so that this distribution satisfies

$$\int_0^\infty d\eta\, \eta^{-\frac{1}{2}} f(\eta) = 2^{-\frac{1}{2}} \Gamma\left(\frac{\beta+1}{2}\right) \Big/ \Gamma\left(\frac{\beta}{2}+1\right) \qquad (28)$$

in addition to the normalization condition.

Here, we shall confine our consideration to systems which are invariant under time reversibility. This is the case in which the system in the chaotic limit is described by a GOE. In this case, the spacing distribution is given by Eq. (26) with β = 1. The consideration of other symmetry classes can be carried out analogously.

RMT within the sub-extensive regime Tsallis statistics, which has been considered in [10] for the three symmetry universalities, is included in the superstatistical formalism by taking $f(\eta)$ to be the gamma distribution in Eq. (8). With this distribution together with condition (27), we obtain

$$P_{\text{Ts}}(s) = \frac{1}{2}\pi s \left[1 + \frac{\pi(q-1)s^2}{2(3-q)}\right]^{-\frac{1+q}{2(q-1)}}. \tag{29}$$

The spacing distribution that corresponds to a uniform distribution can be obtained by substituting Eq. (10) into Eq. (26). After integration, we obtain

$$P_{\text{Uniform}}(s,b) = \frac{1}{4b\pi s^3}\left[\left\{16\pi + (b-2\pi)^2 s^2\right\}e^{-\frac{(b-2\pi)^2 s^2}{16\pi}} - \left\{16\pi + (b+2\pi)^2 s^2\right\}e^{-\frac{(b+2\pi)^2 s^2}{16\pi}}\right], \tag{30}$$

where the parameter $\eta_0$ is set equal to $(b^2 + 4\pi^2)/8\pi$ in order to satisfy the condition that the mean spacing is unity. The parameter $b$ can be expressed in terms of the non-extensivity $q$ by means of Eq. (12).

Our third choice for $f(\eta)$ is two-level distribution of subsection 3.3. For this distribution, we obtain

$$P_{\text{Two-level}}(b) = \frac{1}{2}s\left[\left(\eta_0 - \frac{b}{2}\right)e^{-\frac{1}{2}\left(\eta_0 - \frac{b}{2}\right)s^2} + \left(\eta_0 + \frac{b}{2}\right)e^{-\frac{1}{2}\left(\eta_0 + \frac{b}{2}\right)s^2}\right]. \tag{31}$$

The parameter $\eta_0$ can easily be expressed in terms of the parameter $b$ by substituting Eq. (13) into Eq. (26) and integrating. Both parameters can then be expressed in terms of the effective non-extensivity using Eq. (14).

Finally, we consider the case when the parameter distribution is given by the modified Gaussian distribution of subsection 3.4. In this case, we obtain

$$P_{\text{Gaussian}}(s) = \frac{s}{2\,\text{erf}(\eta_0/b)}e^{\frac{1}{16}\eta b^2 s^4}\left[\left(\eta_0 - \frac{1}{4}b^2 s^2\right)\left\{1 - \text{erf}\left(\frac{1}{4}bs^2 - \frac{\eta_0}{b}\right)\right\}e^{-\frac{1}{2}bs^2}\right.$$
$$\left.+ \left(\eta_0 + \frac{1}{4}b^2 s^2\right)\left\{1 - \text{erf}\left(\frac{1}{4}bs^2 + \frac{\eta_0}{b}\right)\right\}e^{\frac{1}{2}bs^2}\right]. \tag{32}$$

We determine the parameter $\eta_0$ by substituting Eq. (15) into Eq. (26) and integrating, which yields

$$\eta_0 = x^3 e^{-x^2}\left[\frac{\pi_0 F_1(5/4; x^4/16)}{2\Gamma(5/4)\text{erf}(x)}\right]^2, \text{ with } x = \eta_0/b, \tag{33}$$

where $_0F_1(a; z)$ is a confluent hypergeometric function [22]. The parameter $x$ can directly be expressed in terms of the effective non-extensivity $q$ by means of Eq. (16).

We have calculated the NNS distributions for the ensembles under consideration for the four superstatistical ensembles under consideration. For this purpose, we use Eqs. (27) and (29)-(32) with effective non-extensivity $q = 1$ (GOE), 1.1 and 1.3. The results of calculation are shown in Fig. 3. The figure demonstrates the gradual departure from the GOE distribution as the fluctuation of the ensemble parameter increases. However, there is an upper limit beyond which one cannot increase $q$ in the case of the gamma, uniform and two-level superstatistics. These upper limits are defined by Eqs. (10), (12) and (14), respectively. The modified Gaussian distribution of subsection 3.4 does not put any restriction on the value of $q$. However, as $q$ exceeds a value of 1.3 the shape of the spacing distribution remains practically unchanged. The "limiting" spacing distributions of the four versions of $f(\eta)$ are compared in Fig. 4 with the Wigner and Poisson distributions. Although these limiting distributions do not exactly have the same shape, they have their peaks nearly in the same position on the scale of level spacing, at about one third of the distance from the peak of the Wigner distribution to that of the Poissonian (the origin).

**6. Summary and Conclusion**

In this article, we apply superstatistics introduced by Beck and Cohen to describe mixed systems within the framework of RMT. The matrix-element distribution is expressed as a weighted average over the corresponding distribution obtained by the standard RMT with respect to the distribution parameter, i.e. the inverse variance of the matrix elements. The proposed generalization preserves basis independence on the expense of violating matrix-element independence, in variance with most of random-matrix models of mixed systems. We have derived analytical expressions for the joint matrix-element distribution, the level density, and the NNS distribution for mixed systems at least in the first stage of transition out of chaos. The results obtained here are for four different distributions of the inverse variance of matrix elements; one of them leads to Tsallis' non-extensive statistics. We follow the evolution trends of the level density and NNS distribution for each of the four superstatistics in terms of a single parameter, namely the effective non-extensivity $q$, which is defined in terms of the relative dispersion of the ensemble parameter. The results obtained for the four superstatistics agree with each others for values of $q$ in the range of $1 < q \lesssim 1.2$, which suggests that $q$ may serve as a control parameter at least for the early stages of transition from chaos to order.

The four superstatistical NNS distributions are explicitly calculated for systems with time reversal invariance. They undergo a gradual transition from the Wigner form towards the Poissonian but they stop their evolution in the midway of the transition. The termination of the transition out of chaos happens even in the case of a Gaussian parameter distribution, although it imposes no limit on increasing the effective non-extensivity. Accordingly, the proposed superstatistical approach, which preserves base invariance of the random-matrix ensemble, cannot describe systems close to integrality. In fact, it is hard to believe that nearly integrable systems would be modeled as a basis-invariant random ensemble, even if they could be described by a random matrix Hamiltonian.

**References**

[1] O. Bohigas, M.J. Giannoni, C. Schmidt, Phys. Rev. Lett. 52 (1984) 1184.
[2] M.L. Mehta, Random Matrices, 2$^{nd}$ Edition, Academic Press, New York, 1991.
[3] T. Guhr, A. Müller-Groeling, H.A. Weidenmüller, Phys. Rep. 299 (1998) 189.
[4] N. Rosenzweig, C.E. Porter, Phys. Rev. 120 (1960) 169.
[5] F. Haake, Quantum Signatures of Chaos, Springer, Heidelberg, 1991; V.K.B. Kota, Phys. Rep. 347 (2001) 223.
[6] M.S. Hussein, M.P. Sato, Phys. Rev. Lett. 70 (1993) 1089.
[7] C. Tsallis, J. Stat. Phys. 52 (1988) 479.
[8] R. Ballian, Nuovo Cim. 57 (1958) 183.
[9] F. Toscano, R.O. Vallejos, C. Tsallis, Phys. Rev. E 60 (2004) 066131.
[10] A.Y. Abul-Magd, Phys. Lett. A 333 (2004) 16.
[11] C. Beck, E.G.D. Cohen, Physica A 322 (2003) 267.
[12] C. Beck, Europhys. Lett. 64 (2003) 151; A. Reynolds, Phys. Rev. Lett. 91 (2004) 084503; C. Beck, Physica D 193 (2004) 159; K.E. Daniels, C. Beck, E. Bodenschatz, Physica D 193 (2004) 208.

[13] F. Sattin, L. Salasnich, Phys. Rev. E 65 (2002) 035106(R).
[14] C. Beck, Physica A 331 (2004) 173.
[15] M. Ausloos, K. Ivanova, Phys. Rev. E 68 (2003) 046122.
[16] G. Wilk, Z. Wlodarczyk, Phys. Rev. Lett. 84 (2000) 2770.
[17] C. Beck, Phys. Rev. Lett. 87 (2001) 180601.
[18] A. Y. Abul-Magd, H. L. Harney, M. H. Simbel, H. A. Weidenmüller, Phys. Lett. B 597 (2004) 278.
[19] A. Y. Abul-Magd, H. L. Harney, M. H. Simbel, H. A. Weidenmüller, e-print physics/0212049; Ann. Phys. (N.Y.) to be published.
[20] C.E. Porter, Statistical Theory of Spectra: Fluctuations, Academic Press, New York, 1965.
[21] S. Wolfram, The Mathematica Book, 4$^{th}$ ed., Cambridge University Press, Cambridge, 1999.
[22] M. Abramowitz, I.A. Stegun, Handbook of Mathematical Functions, Dover, New York, 1965.
[23] M. Jimbo, T. Miwa, Y. Mori, M. Sato, Physica D 1(1980) 80.
[24] V. K. B. Kota, S. Sumedha, Phys. Re. E 60 (1999) 3045.
[21] P. Chau Huu-Tai, N.A. Smirnova, P. Van Isacker, J. Phys. A 35 (2002) L199.

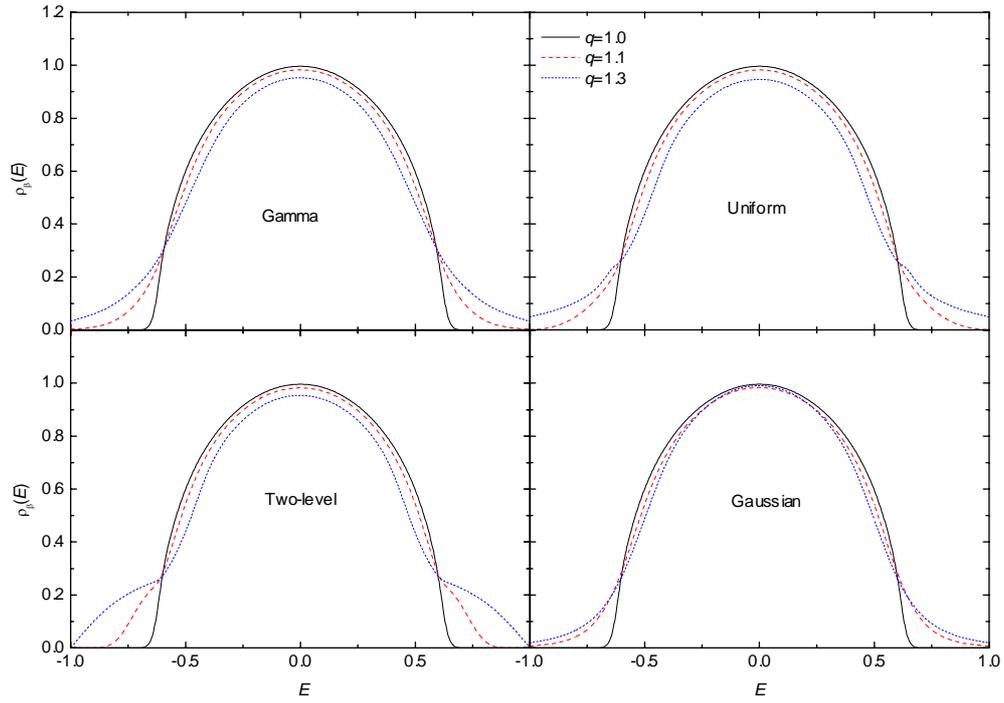

Fig. 1. Level density for a superstatistical ensembles having a Gaussian, uniform and two-level parameter distributions, calculated for different effective non-extensivity parameter $q$. The parameter distributions are centered at the value of $\eta_0 = (\pi/2)^2 \beta/N$ that corresponds to a unit central density for the corresponding Gaussian random ensemble.

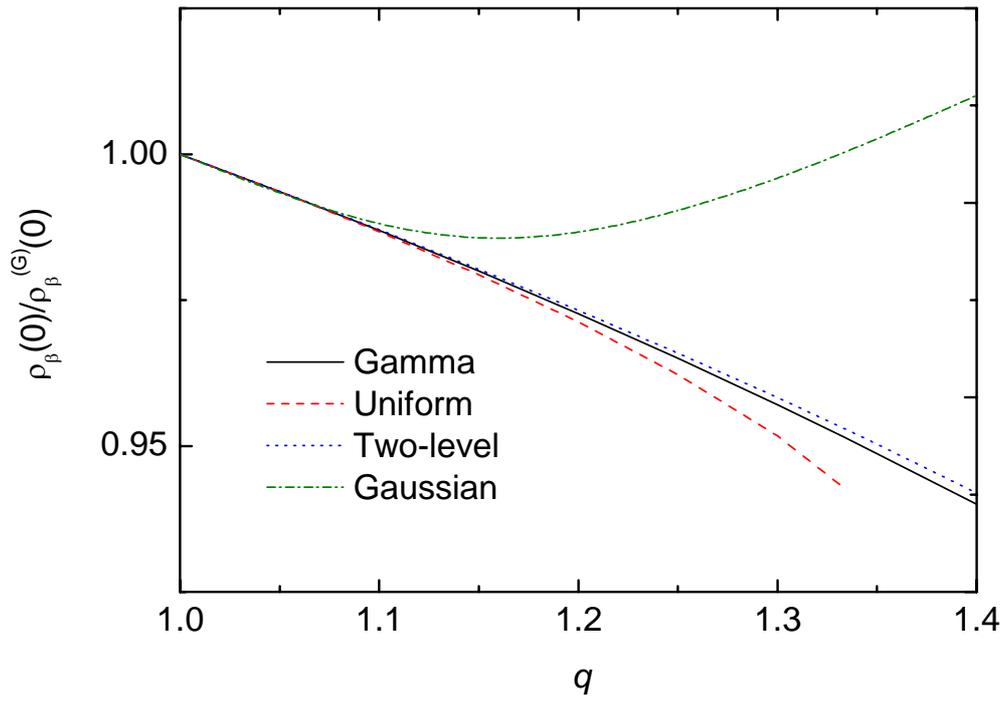

Fig. 2. Dependence of the ratio of the central level densities for superstatistical ensembles with gamma, uniform, two-level and modified-Gaussian parameter distribution to the central level density of the corresponding Gaussian random ensemble on the effective non-extensivity parameter $q$.

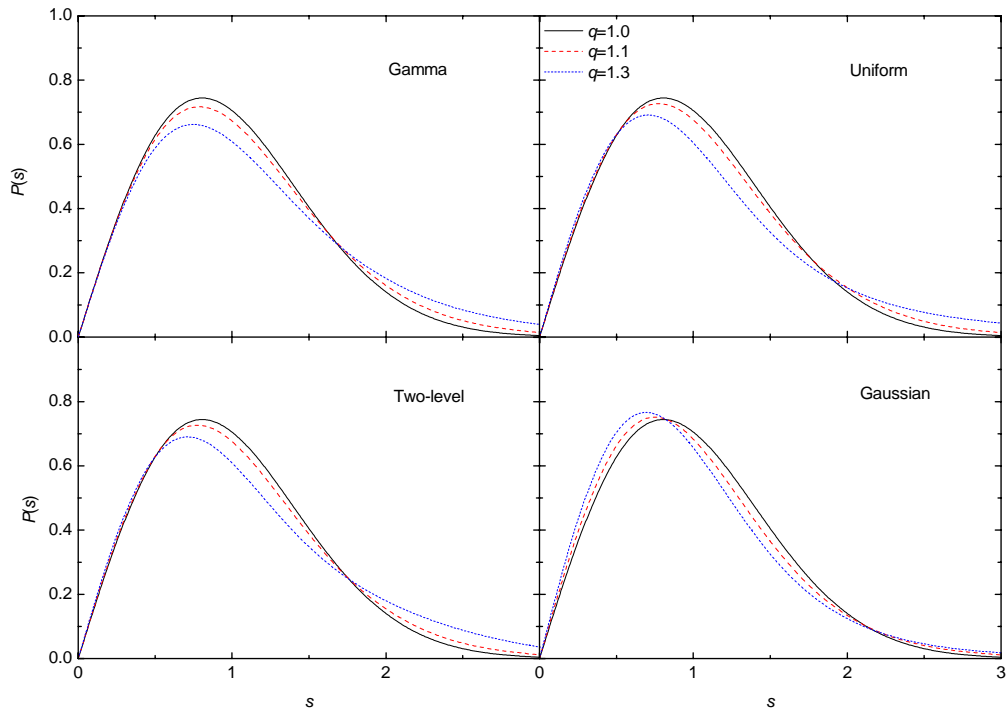

Fig. 3. NNS distributions for superstatistical ensembles having a gamma, uniform, two-level and modified-Gaussian parameter distributions, calculated for different effective non-extensivity parameter $q$.

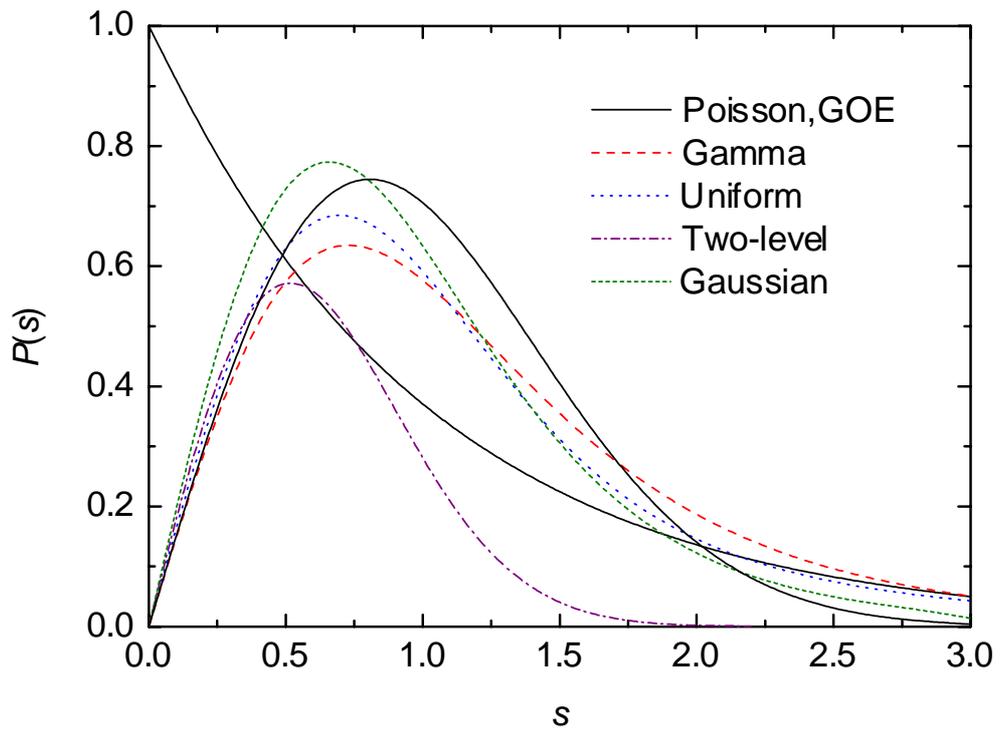

Fig. 4. The limiting distributions of NNS for the four superstatistics under consideration compared to the Wigner and Poisson distributions.